\documentclass[prl,amsmath,amssymb,twocolumn,superscriptaddress,showpacs]{revtex4}

\usepackage{amsmath,amssymb}
\usepackage[usenames]{color}
\usepackage{amssymb}
\usepackage{grffile}
\usepackage[pdftex]{graphicx}
\usepackage{amsmath, amstext, amssymb, amsfonts, amsxtra}
\usepackage{textcomp}
\usepackage{xspace}

\def \ell{{d}}

\newcommand{\hannover}{Institut f\"ur Theoretische Physik, Leibniz Universit\"at Hannover, Appelstr. 2, DE-30167 Hannover, Germany}

\begin{document}

\title{Anyon Hubbard Model in One-Dimensional Optical Lattices}                     

\author{Sebastian Greschner} 
\affiliation{\hannover}
\author{Luis Santos} 
\affiliation{\hannover}

\begin{abstract}
Raman-assisted hopping may be used to realize the anyon Hubbard model in one-dimensional optical lattices. 
We propose a feasible scenario that significantly improves the proposal of [T. Keilmann et al., Nature Commun. 2, 361 (2011)], allowing as  
well for an exact realization of the two-body hard-core constraint, and for controllable effective interactions without the need of Feshbach resonances.
We show that the combination of anyonic statistics and two-body hard-core constraint  leads to a rich ground state physics,  
including Mott insulators with attractive interactions, pair superfluids, dimer phases, and multicritical points. Moreover, 
the anyonic statistics results in a novel two-component superfluid of holon and doublon dimers, characterized by a large but finite compressibility and 
a multipeaked momentum distribution, which may be easily revealed experimentally.
\end{abstract}



\maketitle



Particles are classified as bosons or fermions depending on whether their wave function is symmetric or antisymmetric under exchange. 
Other types of quantum statistics are, however, possible in lower dimensions. Remarkably, 2D systems allow for the existence of anyons, 
i.e. particles with fractional statistics interpolating between bosons and fermions~\cite{Leinaas1977,Wilczek1982, Canright1990}. 
Anyons play a fundamental role in key areas of modern physics, as fractional quantum Hall effect~\cite{Laughlin1983,Halperin1984,Camino2005,Kim2005} and topological quantum computing~\cite{Nayak2008}. Fractional statistics is, however, 
not exclusive of 2D systems~\cite{Haldane1991b}. In particular, 1D anyons have attracted a large deal of
interest~\cite{Haldane1991a, Ha1994, Murthy1994, Wu1995, Zhu1996, Amico1998,Kundu1999, Batchelor2006, Girardeau2006, 
Calabrese2007, DelCampo2008, Hao2009, Hao2012, Wang2014}, although the experimental realization of a 1D anyon gas is still lacking.

Ultracold atoms offer extraordinary possibilities for the analysis of interesting 
many-body systems~\cite{Bloch2008}. In particular, several ideas have been proposed for the 
creation and manipulation of anyons in cold gases~\cite{Paredes2001, Duan2003, Micheli2006,Aguado2008,Jiang2008}. 
Particularly interesting is the recent proposal for the realization of the anyon Hubbard model~(AHM) using Raman-assisted hopping 
in 1D optical lattices~\cite{Keilmann2011}. In this proposal the anyonic statistics may be controlled at will, opening the possibility for the observation 
of statistically induced quantum phase transitions~\cite{Keilmann2011}, asymmetric momentum distributions~\cite{Hao2009}, and intriguing 
particle dynamics in the lattice~\cite{DelCampo2008,Hao2012,Wang2014}. 

In this Letter we first discuss a scheme for realizing the AHM that, although following Ref.~\cite{Keilmann2011}, solves 
crucial drawbacks that would render the original proposal, in general, unfeasible.  The scheme also allows for controllable effective interactions without the need of Feshbach resonances,  
and for an  exact two-body hard-core constraint~(2BHCC), contrary to the approximate realization of 2BHCC resulting 
from Zeno projection due to large three-body loss rates~\cite{Daley2009}, where non-negligible losses are typically present. 
Neither the controllable interactions nor the inherent 2BHCC were considered in Ref.~\cite{Keilmann2011}, which
focused exclusively on statistically induced superfluid~(SF) to Mott insulator~(MI) transitions. We show that the interplay of anyonic statistics, 
2BHCC, and controllable interactions, results in a far richer physics for the AHM, including pair superfluid~(PSF), a dimer~(D) phase, and 
an exotic partially paired~(PP) phase. The latter constitutes, to the best of our knowledge, a novel two-component superfluid 
characterized by a large, but finite, compressibility and a multipeaked momentum distribution, which may be readily revealed in time-of-flight~(TOF) experiments.

 \begin{figure}[t]
 \vspace*{-0.3cm}
 \begin{center}
 \includegraphics[width =\columnwidth]{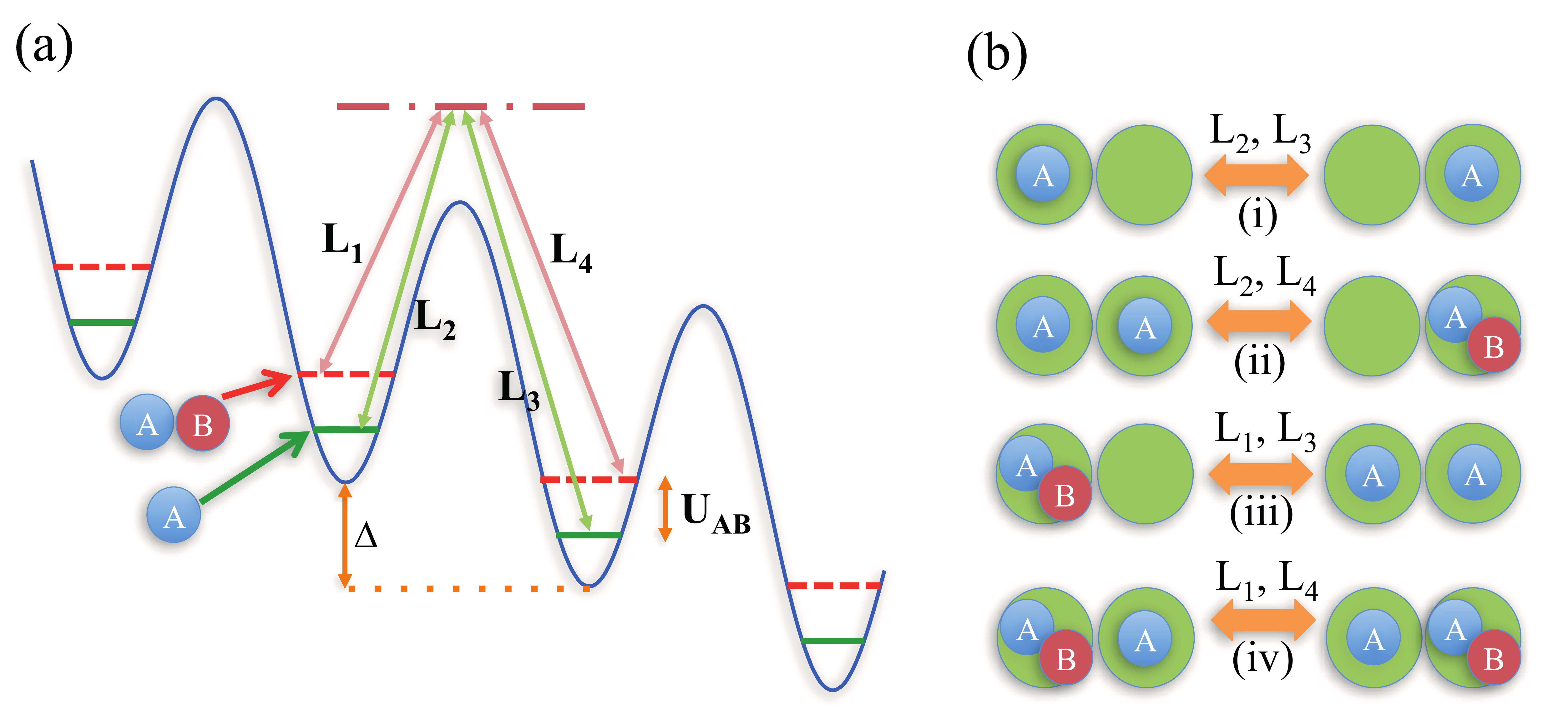}
 \caption{(a) Raman scheme proposed for the realization of the AHM; (b) Raman assisted hops (i)--(iv) discussed in the text.}
 \vspace*{-0.8cm}
 \label{fig:1}
 \end{center}
 \end{figure}



\begin{figure*}[ht]
\centering
\includegraphics[width=1.0\textwidth]{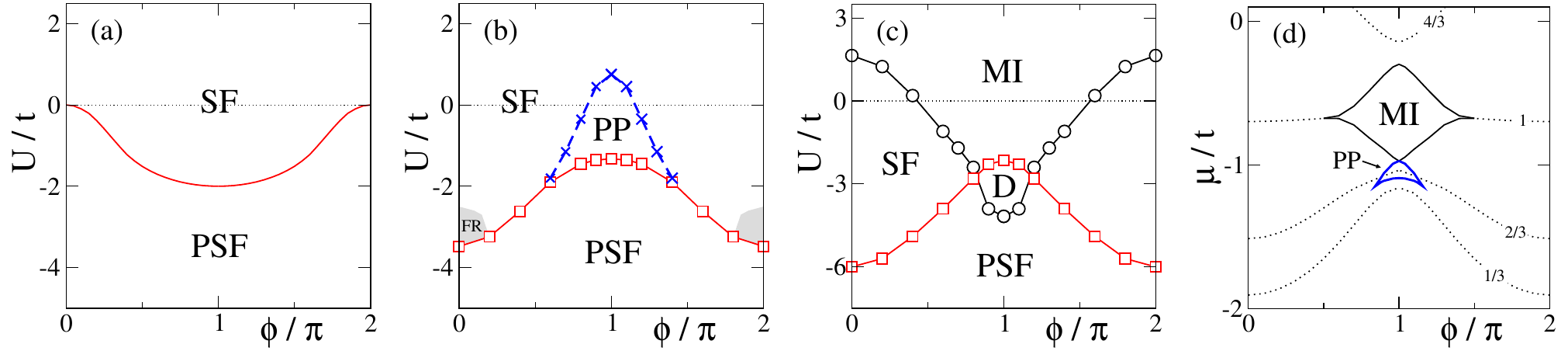} 
\caption{Phase diagram of the AHM: (a) dilute limit ($\rho\to0$), (b) incommensurate intermediate fillings (here $\rho=1/2$) and (c) unit filling ($\rho=1$). The PSF-SF transition lines~(boxes or solid line in (a)), extracted from the crossing of two and one particle excitations coincides with the onset of parity order. At incommensurate fillings for small $\phi$ the system exhibits a first order transition from SF to PSF phase through a forbidden region (FR, dotted area). For $\phi / \pi\gtrsim 0.6$ we find the PP phase. For unit filling the Ising-type PSF-SF transition and the 
Kosterlitz-Thouless SF-MI transition lines cross around $\phi / \pi\approx0.8$ in a multicritical point. For larger values of $\phi$ the system exhibits a dimerized gapped phase D. (d) Grand-canonical phase diagram and lines of constant density $\rho$ for $U=0$ as function of $\mu$ and $\phi$. Solid lines denote the MI and PP phase, the rest is SF. For $\mu/t<-2$ ($\mu/t>4$) the vacuum $\rho=0$ (fully occupied phase, $\rho=2$) is realized.}
\label{fig:gs_pd_n}
\end{figure*}

\emph{Anyon-Hubbard model.--} 
We briefly describe our proposal for the realization of the AHM~\cite{footnote-SM}.
We consider atoms, bosons or fermions, with states $|A\rangle$ and $|B\rangle$, in a deep tilted spin-independent 1D lattice with no direct hopping. 
For the specific case of $^{87}$Rb, the choice would be 
$|A\rangle\equiv |F=1,m_F=-1\rangle$ and $|B\rangle\equiv |F=2,m_F=-2\rangle$. 
Both states are coupled far from resonance by D$_1$ lasers $L_{1,4}$~(with linear polarization) and $L_{2,3}$~(with circular $\sigma_-$ polarization)~(Fig.~\ref{fig:1}(a)). 
Due to selection rules, $|B\rangle$ is just affected by lasers $L_{1,4}$. In contrast, both $L_{2,3}$ and $L_{1,4}$ couple with $|A\rangle$, but the coupling with $L_{1,4}$ 
can be made much smaller than that of $L_{2,3}$, and does not affect the main conclusions of this Letter~\cite{footnote-spurious}. We hence assume below that $|A\rangle$ is just affected by $L_{1,2}$.
These lasers, with frequencies $\omega_{j=1,\dots,4}$, induce four Raman-assisted hops~(Fig.~\ref{fig:1}(b)):
(i) $(A,0)\rightarrow(0,A)$, given by $L_{2,3}$, such that $\omega_2-\omega_3=-\Delta$~(with $\Delta$ the lattice tilting);  
(ii) $(A,A)\rightarrow(0,AB)$, assisted by $L_{2,4}$ with $\omega_2-\omega_4=-\Delta+U_{AB}+U$;   
(iii) $(AB,0)\rightarrow(A,A)$, given by $L_{1,3}$ with $\omega_1-\omega_3= -\Delta-U_{AB}-U$;   
(iv) $(AB,A)\rightarrow(A,AB)$, assisted by $L_{1,4}$, such that $\omega_1-\omega_4=-\Delta$.  
The notation $(\eta_L,\eta_R)$ denotes the state of neighboring sites $L$ and $R$: $\eta_{L,R}=0$ (vacuum), $A$~(single $A$ occupation) 
or $AB$~(an $A$ and a $B$ particle);  $U_{AB}$ characterizes the $AB$ interaction. The frequency differences compensate 
the tilting, avoiding Bloch oscillations; moreover, the detuning $U$ acts as a controllable effective on-site interaction, as shown below.

We assume the width of the Raman resonances, $W$, is small-enough such that each Raman process may be addressed independently. If the atoms are fermions, the only undesired process 
$(A,0)\rightarrow (0,B)$~(where $B$ denotes a site with a single $B$ atom) 
may be avoided if $U_{AB}\gg W$. For bosons, $(A,A)\rightarrow (0,AA)$ and $(AB,A)\rightarrow (B,AA)$~(with $AA$ a site with two $A$ atoms) must be also 
avoided. This demands $U_{AA}-U_{AB}$, $U_{AA}$, $U_{AB} \gg W$, where $U_{AA}$ characterizes $AA$ interactions~\cite{footnote-SM}. 

The laser $L_j$ is characterized by a Rabi frequency $\Omega_j e^{i \phi_j}$, 
with $\phi_1=-\phi$, and $\phi_{j\neq 1}=0$.  We impose 
$\frac{|\Omega_1||\Omega_4|}{4}=\frac{|\Omega_2||\Omega_3|}{3}=
\frac{|\Omega_1||\Omega_3|}{2\sqrt{3}}=\frac{|\Omega_2||\Omega_4|}{2\sqrt{3}}=\Omega^2$ (for $^{87}$Rb; for other species Clebsch-Gordan coefficients may differ~\cite{footnote-SM}). 
$|\Omega_j|$ is chosen such that they mimic bosonic enhancement, absent otherwise due to the distinguishability between $A$ and $B$; $\epsilon=1$~($-1$) for bosons~(fermions). 
Under these conditions the system reduces to an effective single-component 1D Bose-Hubbard model~\cite{footnote-SM}, with on-site Fock states 
$\{|0\rangle, |1\rangle\equiv |A\rangle, |2\rangle\equiv |AB\rangle\}$, and an occupation-dependent Peierls phase~\cite{Keilmann2011,footnote-SM}:

\begin{eqnarray}
\!\!\!\!\!\! H \! &=& \!  - t \sum_j (b_{j}^{\dagger} {\rm e}^{i\phi\, n_j} b_{j+1}^{\phantom \dagger} + \text{H.c.}) +
\frac{U}{2} \sum_j  n_j (n_j-1)
\label{eq:ham}
\end{eqnarray}
where $U$ is an effective coupling constant whose sign and magnitude may be tailored 
without the need of Feshbach resonances by properly choosing the laser frequencies;  
$t\simeq J(\Omega^2/\delta)/2\Delta$ is the effective hopping rate~\cite{MiyakeThesis}, with $J$ the lattice hopping in the absence of tilting, and $\delta\gg U_{AB}, \Delta, \Omega$ the detuning from the single-photon transitions; 
$b_j^{\dagger}$~($b_j^{\phantom \dagger}$) are creation~(annihilation) operators at site $j$, and $n_j=b_j^{\dagger}b_j^{\phantom \dagger}$.
Introducing $\alpha_j\equiv {\rm e}^{i \phi \sum_{1\leq l\leq j-1} n_l} b_j$, one obtains the anyon-Hubbard model~(AHM)~\cite{Keilmann2011}:
\begin{eqnarray}
H \! &=& \!  - t \sum_j (\alpha_{j}^{\dagger}\alpha_{j+1}^{\phantom \dagger} + \text{H.c.}) + \frac{U}{2} \sum_j  n_j (n_j-1)
\label{eq:ham_anyons}
\end{eqnarray}
where $\alpha_j,\alpha_j^\dagger$ satisfy anyonic commutations~\cite{Fradkin1991}: $\alpha_j \alpha_k^\dagger - \rm{e}^{-i\phi\, {\rm sgn}(j-k)} \alpha_k^\dagger \alpha_j = \delta_{jk}$ and $\alpha_j \alpha_k - \rm{e}^{-i\phi\, {\rm sgn}(j-k)} \alpha_k \alpha_j = 0$.

None of the Raman-assisted hoppings leads to triple occupation, resulting in an effective 2BHCC, $(b_j^\dag)^3=0$. 
This inherent constraint  turns out to be crucial. It prevents collapse for a large $U/t<0$, leading to a rich physics of quantum phases~(see Fig.~\ref{fig:gs_pd_n})~\cite{footnote-numerics}.

Note that although hops (i) and (iv) share the same energy difference $-\Delta$, they may be independently addressed, even if 
$\delta\gg U_{AB},\Delta$.  
This point is crucial, solving the major drawback of Ref.~\cite{Keilmann2011} in which only one ground state component was considered, and hence 
hops (i) and (iv) cannot be independently addressed unless $\delta\ll U_{AB},\Delta$, which for typical experimental parameters would lead to 
large heating. An exception may be given by the use of clock transitions in alkaline-earth atoms~\cite{Cazalilla2014}); however, 
the adiabatic elimination of the excited state demands $\Omega\ll\delta$, and hence $t\ll\delta\ll U_{AB},\Delta$, 
would lead to very long experimental time scales~\cite{footnote-Scales}.


\begin{figure*}[ht]
\centering
\includegraphics[width=1.0\textwidth]{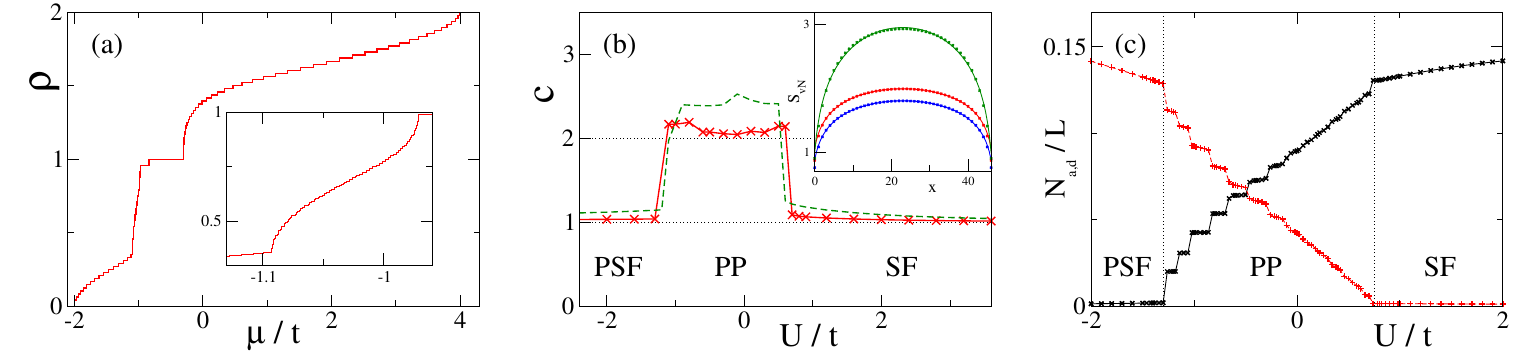} 
\caption{Phase transitions for $\phi=\pi$. (a) Equation of state, density $\rho$ as function of chemical potential $\mu$ 
 (calculated for $L=48$ sites, $U=0$). The inset shows a detailed view on the PP region for $L=240$ sites. (b) Estimation of the central charge for $L=24$ (dashed line) and $L=48$ (solid line) sites and periodic boundary conditions ($\rho=1/2$). The inset shows fits to the block-entanglement entropy $S_{vN}$ for (from bottom to top) $U=2, -2, -1.0$. (c) Sum of local single-~($N_a$) and two-particle correlations~($N_d$) for $L=60$ sites and $\rho=1/2$, see text.}
\label{fig:cut_pd_n0.5}
\end{figure*}

\emph{Effective repulsion.--} The dilute limit~(lattice filling $\rho\ll 1$) is best understood from the analysis of the two-particle problem~\cite{Kolezhuk2012}, characterized by the scattering length (in units of the lattice spacing) 
\begin{equation}
a = \frac{-(1+\cos\phi)}{4(1-\cos\phi)+2U/t}
\label{eq:aAnyon}
\end{equation}
By comparison to a 1D gas of bosons with mass $m$ one extracts an effective interaction strength $g=-2/(a m)$ ~\cite{Kolezhuk2012}. 
Thus the AHM behaves as a repulsively interacting system for small $\rho$ and $U\to 0$, since even in that limit $g$ remains finite and positive for any $\phi>0$, approaching 
Tonks limit as $\phi\to\pi$.
This is consistent with a weak coupling analysis~\cite{Greschner2014}, in which the Luttinger liquid~(LL) parameter 
$K=\pi/(\phi^2+\frac{U}{2\rho t})^{1/2}$, showing that $\phi$ has qualitatively the same effect as a repulsive $U>0$.
In Fig.~\ref{fig:gs_pd_n}(a) we depict the phases for $\rho\to 0$~\cite{footnote-boundstate}: a SF phase, and a PSF of bound pairs at $U<0$. 
Due to the effective repulsion the SF extends into the $U<0$ region for $\phi>0$.

\emph{Incommensurate fillings.--} Bose models~($\phi=0$) with 2BHCC have been recently studied in different scenarios~\cite{Daley2009, Bonnes2011, Greschner2013}. 
A SF to MI transition results for $\rho=1$ and large-enough $U>0$; for $U<0$ a transition to a PSF occurs, 
which at incommensurate $\rho$ may become first order~\cite{footnote-firstorder}.
This forbidden region~(FR) occupies the shaded area in Fig.~\ref{fig:gs_pd_n}(b). 
At fixed $\rho$ the ground state is characterized by the formation of a macroscopically bound state with a phase separation of doublons and holes. 
The FR shrinks, however, fast with increasing $\phi$.

For $\phi \gtrsim 0.6 \pi$ a novel gapless phase, henceforth called partially paired~(PP) phase, appears between the PSF and SF phases. As depicted in Fig.~\ref{fig:cut_pd_n0.5}(a), a sharp kink in the $\rho(\mu)$ curve~(at $\mu\simeq -t$) accompanies the onset of the PP phase. 
Whereas SF~(PSF) is characterized by steps of $1$~($2$) particles in the $\rho(\mu)$ curve, the PP phase exhibits a complex sequence of steps of $2$ and $1$ particles. 
Even more relevant is the behavior of the central charge $c$, which we calculated from the conformal expression of the von Neumann entropy,  
$S_{vN, L}(l) = \frac{c}{3} \ln\left[ \frac{L}{\pi} \sin\left(\frac{\pi}{L}l\right) \right] + \gamma$, for a subsystem of length $l$ in a system of $L$ sites, with $\gamma$ a constant~\cite{Vidal2003, Calabrese2004}. 
As shown in Fig. ~\ref{fig:cut_pd_n0.5}(b), contrary to the SF and PSF phases~(with $c=1$), the PP phase is a two-component gapless phase characterized by $c=2$.

\emph{Nature of the PP phase.--} The features of the PP phase may be understood from the following simplified picture. 
Due to the 2BHCC the hopping term in Eq.~\eqref{eq:ham} is $J_1 b_j^\dagger b_{j+1} + J_2 b_j^\dagger n_j b_{j+1}$, with $J_1/t = 1$ and $J_2/t=-1 + {\rm e}^{i \phi}$.
For $\phi\approx\pi$ correlated hopping dominates, $J_2=-2 J_1$, and 
doublons, or more precisely dimers, $ \left|d\right>_j = \alpha \left|11\right>_{j,j+1} + \beta \left|20\right>_{j,j+1}$, 
gain a large binding energy $-\sqrt{2} J_2$.
However, dimer hopping, $J_1/\sqrt{2}$, is reduced compared to single particles. 
For a certain range of small $U$ and $\rho$ it is energetically favorable to occupy both doublon dimers $\left|d\right>_j$ and atomic dimers,  
$ \left|a\right>_j = \tilde\alpha \left|01\right>_{j,j+1} + \tilde\beta \left|10\right>_{j,j+1}$. 
Neglecting interactions between these two quasiparticles, reasonable for small doublon and atom densities, $\rho_d$ and $\rho_a$, with $\rho_a+\rho_d\ll 1$, 
one arrives at an effective model with two independent hardcore bosons, $H_{ad}=H_a+H_d$, with~(for $U=0$)
\begin{eqnarray}
H_a &=& J_1 \sum_j (a_j^\dagger a_{j+1}+{\mathrm H. c.})- J_1 \sum_j a_j^\dagger a_{j}, \\
H_d &=& - \frac{J_1}{\sqrt{2}}e^{i\arg J_2} \sum_j d_j^\dagger d_{j+1}- \sqrt{2} J_2 \sum_j d_j^\dagger d_{j} 
\label{eq:ham_ad}
\end{eqnarray}
where $a_j$ ($a_j^\dagger$) and $d_j$ ($d_j^\dagger$) are annihilation~(creation) operators of hardcore atom and doublon dimers on sites $j,j+1$, 
and $\rho_a + 2\rho_d = \rho$. In spite of being a gross oversimplification, $H_{ad}$ captures the qualitative features of the PP phase. 
For $U\approx 0$, at low $\rho$ the ground state only contains $a$ dimers, whereas for higher $\rho$~($\gtrsim0.3$ for $U=0$) both dimers are present. 
The resulting $\rho(\mu)$ curve obtained from $H_{ad}$~(not shown) exhibits an irregular pattern of steps of $2$ or $1$ particles. 
The two-component phase extends for values around $U=0$, whereas for $U\gg J_1$~($U\ll -J_1$) only $a$~($d$) dimers are present~\cite{footnote-transition}. 

The atom and doublon dimerizations, $N_a = \sum_i \langle b_i^\dagger b_{i+1}\rangle$, and $N_d = \sum_i \langle (b_i^\dagger)^2 (b_{i+1})^2\rangle$, 
approximately measure $\rho_a$ and $\rho_d$~(Fig.~\ref{fig:cut_pd_n0.5} (c)). Whereas the SF and PSF phases are LLs 
of almost hardcore single particles or pairs (for $\phi=\pi$), in the PP phase both components coexist. Note the existence of many sharp jumps to states with approximately finite doublon density.

\begin{figure}[t]
\centering
\includegraphics[width=1.0\columnwidth]{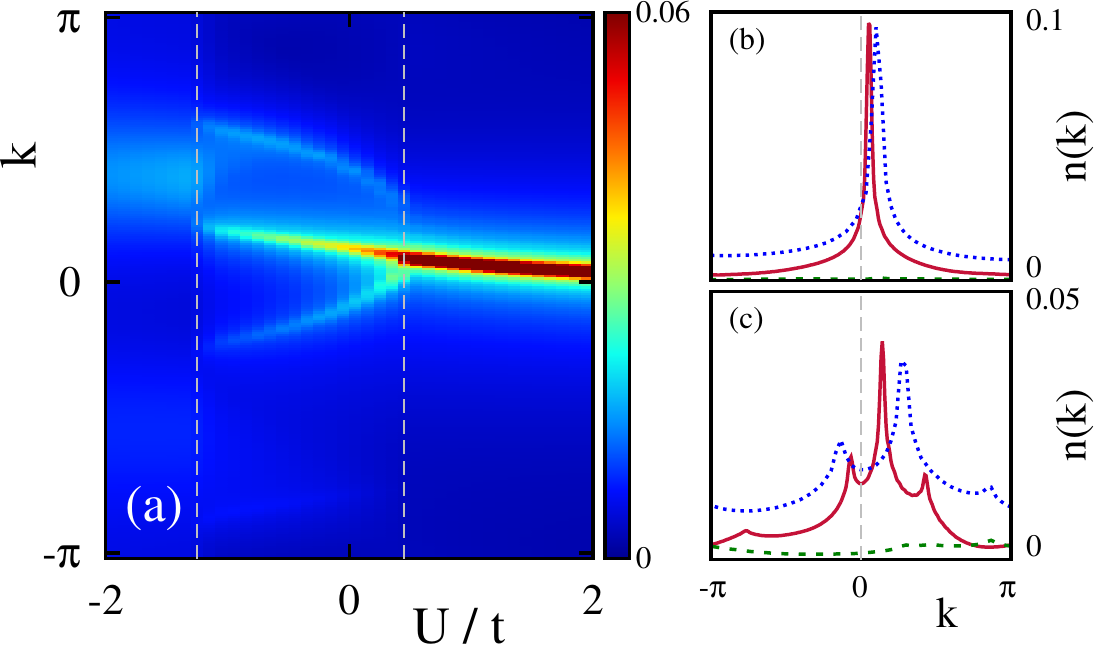} 
\caption{(a) Quasimomentum distribution $n(k)$ for $\rho=1/2$ as
function of $U/t$ for $\phi=0.9 \pi$ ($L=60$ sites). Dashed lines denote phase
transitions between PSF, PP and SF phases. (b) and (c) show the momentum
distribution for $U=2t$ and $U=0$ for the AHM (solid line) and the
components $A$ (dashed line) and $B$ (dotted line) separately (for bosonic atoms). 
The latter are obtained by direct simulation of the two-component model with Raman-induced hopping~(see Suppl. Material for a discussion of the fermionic case).}
\label{fig:mom}
\end{figure}

The AHM is characterized by a density- and $U$-dependent drift of the quasimomentum distribution~\cite{Keilmann2011, Greschner2014}, evident in Fig.~\ref{fig:mom} 
for the SF. Interestingly, whereas the SF presents an usual single-peaked distribution, the PP phase shows a multipeaked distribution. This is because the $a$ and $d$ hop with a different Peierls phase, and hence the overall ($a+d$) 
momentum distribution presents multiple peaks. For PSF, single-particle correlations decay exponentially, and the peak structure blurs.

\emph{Unit filling.--} We consider now the special case of $\rho=1$. Tuning $\phi$ may induce a SF to MI transition~\cite{Keilmann2011, Greschner2014}. This is so even  
for $U<0$~(Fig.~\ref{fig:gs_pd_n} (c)) due to the effective on-site repulsion discussed above. 
This Kosterlitz-Thouless transition may be determined as the curve at which $K=2$~\cite{GiamarchiBook},  
which is consistent with the opening of the energy gap.
As mentioned above, for $U<0$ an Ising-type transition occurs from SF to PSF. 
However, in the vicinity of $\phi\approx \pi$, the combination of anyonic statistics and 
2BHCC results in a dimerized phase, D. The latter is 
characterized by a finite dimer-order parameter $O_D=\left<T_{L/2} - T_{L/2+1}\right>$ with $T_i = b_{i}^\dagger b_{i+1} + H.c.$.
The Ising transition line is determined by the crossing of single- and two-particle excitations, and by a sharp peak in the fidelity susceptibility~\cite{Greschner2013b}.  
Interestingly, the phase diagram presents a SF-MI-D-PSF multicritical point.

\emph{Probing the anyon-Hubbard model--} 
A discussion on experimental detection demands a grand-canonical analysis since local density arguments translate the dependence 
on the chemical potential, $\mu$, into a spatial dependence of the phases in the presence of an overall harmonic confinement. 
Crucially, the PP phase has a finite extension in both $\phi$ and $\mu$~(Fig.~\ref{fig:gs_pd_n} (d)). 
Hubbard models with correlated hopping~(both for fermions~\cite{Aligia2000,DiLiberto2014} and bosons~\cite{Rapp2012}) 
may present a similar (but insulating) phase with coexistent doublons and holons, which is, however, infinitely compressible, i.e. it is a single point as a function of $\mu$. 
As a result these phases cannot be observed in trapped experiments, just leading to abrupt density jumps. On the contrary the novel PP phase, having a finite $\mu$ extension, may be experimentally observable by properly setting the central $\mu$ in the PP region (adjusting the particle number). 
Moreover, the multipeaked momentum distribution of the PP phase can be easily probed in TOF measurements. 
Figure~\ref{fig:mom}(c) shows that TOF images of A and the B may be employed to reveal the PP phase~\cite{footnote-SM}.

\emph{Summary.--} We propose a Raman scheme that realizes the AHM using cold atoms in 1D lattices, overcoming the main limitations of the proposal of Ref.~\cite{Keilmann2011}. This scheme 
results naturally in a 2BHCC, and allows for tailoring effective on-site interactions without Feshbach resonances. 
The combination of anyonic character and 2BHCC leads to a rich ground state physics, including Mott insulators for attractive interactions, 
a dimerized gapped phase. pair superfluids, and a novel two-component superfluid (partially paired phase) characterized by a large compressibility and a 
peculiar multipeaked momentum distribution that may be easily probed experimentally. Finally, we would like to mention that the Raman scheme 
may be extended to 2D lattices. Although for 2D lattices the boson-anyon mapping is lost, a 2D setup would open the exciting possibility 
of inducing a density-dependent 
magnetic field, contrary to the static field created in recent experiments~\cite{MiyakeThesis,Aidelsburger2013,Miyake2013,Atala2014}. 
This possibility will be investigated in a forthcoming work.

\begin{acknowledgments}
\emph{Acknowledgements.--} We thank A. Eckardt, C. Klempt, D. Poletti,  M. Roncaglia, E. Tiemann, T. Vekua for discussions. We acknowledge support by the 
Cluster QUEST and the DFG Research Training Group 1729. 
Simulations were carried out on the cluster of the Leibniz Universit\"at Hannover.
\end{acknowledgments}

\bibliographystyle{prsty}

\begin{thebibliography}{}
\bibitem{footnote-Scales} Typical experimental values are $U_{AB}$ and $\Delta$ of the order of $100$ Hz to few kHz, $\Omega$ of tens of kHz, $\delta$ of few GHz, and hence $t$ of the order of few kHz. Demanding $\delta\ll U_{AB},\Delta$, hence requires $\Omega\ll\delta\ll$few kHz, and hence $t$ well below $1$ Hz.
\bibitem{footnote-boundstate} While expression~\eqref{eq:aAnyon} correctly describes the effective repulsive behavior for positive and $U<0$, this simplified two-particle scattering picture does not capture the correct SF-PSF transition line for $\phi\to\pi$ and $U<0$, as shown in Fig.~\ref{fig:gs_pd_n}(a).
\bibitem{footnote-transition} The effective model $H_{ad}$ suggests the SF-PP transition to be of commensurate-incommensurate type. This is consistent with the numerically observed cusp-singularity in the $\mu$--$\rho$ curve, as shown in Fig.~\ref{fig:cut_pd_n0.5}(a).
\end{thebibliography}

\end{document}


\title{Supplementary material for "The anyon-Hubbard model in one-dimensional optical lattices"}

\author{S. Greschner and L. Santos}
\affiliation{Institut f\"ur Theoretische Physik, Leibniz Universit\"at Hannover, Appelstr. 2, DE-30167 Hannover, Germany}

\begin{abstract}
In this Supplementary Material we discuss in more extent the creation of density-dependent gauge fields using Raman-assisted hopping, and 
additional details about detection using the quasi-momentum distribution.
\end{abstract}


\maketitle

\section{Creation of density-dependent gauge fields}

\subsection{Two-component system}
\label{subsec:2comp}

We consider two hyperfine states, A and B, of a bosonic or fermionic atom.  
Both components are confined to the lowest band of a tilted 1D optical lattice along $x$, of spacing $D$, and depth $V_0=sE_R$, with $E_R=\hbar^2 \pi^2/2mD$ 
the recoil energy.  Without tilting there is a hopping rate $J$ to nearest neighbors. The lattice tilting induces an energy shift $\Delta$ from site to site. 

We denote as $w(x-j D)$ the Wannier function maximally localized at site $j$. Due to the tilting, it is convenient 
to use Wannier-Stark states. For $J\ll \Delta$, the Wannier-Stark state centered at site $l$ may be approximated as 
$\psi_{l}(x)\simeq w(x-lD)+\frac{J}{\Delta}(w(x-(l+1)D)-w(x-(l-1)D))$.
The 3D on-site wavefunction at site $j$ 
is $\Phi_j({\bf r})=\psi_j(x)\varphi(y,z)$, where $\varphi(y,z)$ is given by the strong transversal confinement. For simplicity 
we assume below $\varphi(y,z)\simeq w(y)w(z)$. 

On-site interactions between $\alpha$ and $\beta$~($\alpha,\beta=$A,B) are 
characterized by the coupling constant 
$U_{\alpha,\beta}=\frac{4\pi\hbar^2a_{\alpha,\beta}}{m}\int d^3{\mathbf r} |w({\mathbf r})|^4$, with $a_{\alpha,\beta}$ the corresponding scattering length. 
For a sufficiently deep lattice, the evaluation of the on-site interactions 
is simplified by means of the harmonic approximation: $\Phi({\bf r})\simeq (\sqrt{\pi} l)^{-3/2}e^{-r^2/l^2}$, 
where $l=D s^{-1/4}/\pi$. Using this approximation we obtain:
$U_{\alpha,\beta}\simeq \pi (2\pi)^{3/2} s^{3/4}\frac{\hbar^2 a_{\alpha,\beta}}{mD^3}$. Of course, for fermions only inter-species 
on-site interactions are possible.


\subsection{Laser transitions}
 \label{subsec:Lasers}
 
Before discussing Raman-assisted hopping, we comment on specific transitions in alkali atoms. 
We analyze first the case of bosons, which we illustrate with the example of $^{87}$Rb (below we comment on other species). 
We choose A$\equiv |F=1,m_F=-1\rangle$ and B$\equiv |F=2,m_F=-2\rangle$. Note that one could in principle swap the definition of A and B. 
The opposite choice would be however problematic, since the scattering lengths fulfill 
$a(|F=1,m_F=-1\rangle,|F=2,m_F=-2\rangle)=a(|F=2,m_F=-2\rangle,|F=2,m_F=-2\rangle)$~\cite{footnote-collision}. Hence swapping A and B would 
lead to $a_{AA}=a_{AB}$, which would not allow to remove some spurious Raman transitions~(see Subsec.~\ref{subsec:Spurious}).
Note also that since we consider maximally stretched states, hyperfine-changing 
collisions are absent. 

\begin{figure}[t]
\begin{center}
\includegraphics[width=1.0\columnwidth]{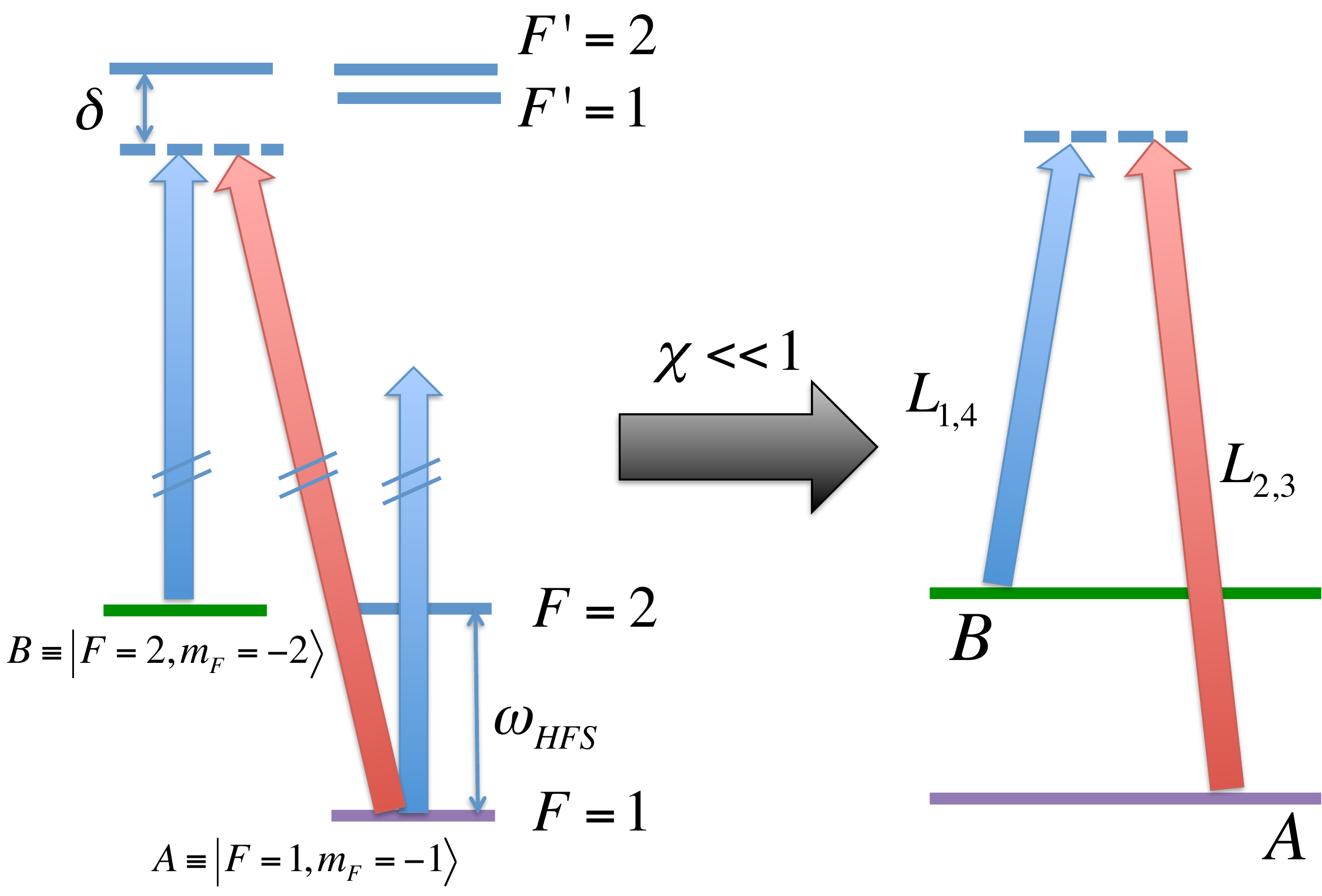}
\vspace{-0.5cm}
\end{center}
\caption{(Color online) Left: Transitions considered for the realization of the Raman-assisted hoppings, with $\pi$ polarized lasers~(blue) and 
$\sigma_-$ polarized lasers~(red); right: if $\chi\ll 1$~(see text), one may assume in a very good approximation that the $\pi$ lasers affect 
just the A state, whereas the $\sigma$ lasers affect just the B state by construction.}
\label{fig:S1}
\end{figure}

We consider two lasers~(see Fig.~\ref{fig:S1}), one with linear polarization ($\pi$ laser), and other with $\sigma_-$ polarization~($\sigma$ laser). 
We define as $\delta$ the detuning of the $\pi$ laser from the $|F=2,m_F=-2\rangle\to |F'=2,m'_F=-2\rangle$ transition, which is~(up to slight detunings of the order of kHz or smaller)
 the same as that of the $\sigma$ laser from the $|F=1,m_F=-1\rangle\to |F'=2,m'_F=-2\rangle$ transition. Both lasers are red detuned. 
Note that due to selection rules $|F=2,m_F=-2\rangle$ is dark with respect to the $\sigma$ laser.
In contrast, the $\pi$ laser links as well $|F=1,m_F=-1\rangle$ to $|F'=2,m'_F=-1\rangle$ and $|F'=1,m'_F=-1\rangle$, with detunings, respectively, $\delta+\omega_{HFS}$ and 
 $\delta+\omega_{HFS}-\omega'_{HFS}$, where $\omega_{HFS}$ and $\omega'_{HFS}$ are the hyperfine splittings of the $^2S_{1/2}$ and $^2P_{1/2}$ manifold, 
 respectively. The fact that the $\pi$ laser affects $|F=1,m_F=-1\rangle$ may introduce spurious effects~(see Subsec.~\ref{subsec:4laser}). 
 
Fermions may be treated in a similar way. For the specific case of $^{40}$K, we choose A$\equiv |F=9/2,m_F=-9/2\rangle$ and B$\equiv |F=7/2,m_F=-7/2\rangle$. 
We may then employ a similar $\pi$--$\sigma$ laser configuration as in Fig.~\ref{fig:S1}. Since $F=9/2$ lies below $F=7/2$ the lasers must be now blue detuned.
The above mentioned choice of A and B is better than the opposite, since it minimizes the spurious effects introduced the the coupling of $|F=7/2,m_F=-7/2\rangle$ by the $\pi$ laser~(see Subsec.~\ref{subsec:4laser}).



 \subsection{Raman-assisted hopping}
 \label{subsec:Raman}

 Direct hopping is prevented for a large-enough tilting, $J\ll\Delta$.
Raman assisted hopping may be induced by the absorption of a  
photon of laser $L_m$ and the emission of a photon of laser $L_n$. 
The amplitude of the Raman-assisted hopping, $J_{nm}$, from site $j$ to site $j+1$, may be 
evaluated following Ref.~\cite{MiyakeThesis}:
\begin{equation}
J_{nm}=\frac{V_{nm}}{4}
\int d^3{\mathbf r}\,  \Phi_{j+1}({\mathbf r})^* e^{i\delta{\mathrm k}^{nm}\cdot {\mathbf r}} \Phi_{j}({\mathbf r})
\end{equation}
where 
$\delta{\mathbf k}^{nm}={\mathbf k}_n-{\mathbf k}_m$ is the difference between the laser wave vectors of the lasers, and $V_{nm}$ is the amplitude of the Raman coupling~(see Subsec.~\ref{subsec:4laser}).
For $J\ll \Delta$:
\begin{eqnarray}
&&\!\!\!\!\!\!\!\!\!\!\!\int dx \psi_{j+1}(x)^* e^{i\delta k_x^{nm}x} \psi_{j}(x)\simeq  \nonumber \\
&&e^{i\delta k_x^{nm}D(j+1/2)}  \left [ \int dx w(x-D)^*w(x) e^{i\delta k_x^{nm}x} \right \delimiter 0 \nonumber \\
&& +\left \delimiter 0 2i\frac{J}{\Delta} \sin\left (\frac{\delta k_x^{nm}D}{2} \right ) \int dx |w(x)|^2 e^{i\delta k_x^{nm}x} \right ].
\end{eqnarray}
For $s\gg 1$~($l\ll D$):
\begin{equation}
J_{nm}\simeq i\left (\frac{V_{nm}}{2\Delta} \right ) J \sin \left ( \frac{\delta k_x^{nm} D}{2} \right ) e^{i \delta k_x^{nm} D(j+1/2)}.
\end{equation}
Note that $\delta k_x^{nm}\neq 0$ is necessary to establish a significant hopping~\cite{MiyakeThesis,MiyakePaper,BlochPaper}.
 
 \subsection{Four-laser scheme}
  \label{subsec:4laser}
 
 We employ the notation of the main text, $(\eta_j,\eta_{j+1})$, to denote the states of neighboring sites, where $\eta_{j,j+1}$ may be $0$~(empty site), 
 $A$ or $B$ (singly occupied site with an A or B component), and $AA$, $AB$ and $BB$ for doubly-occupied sites.

 
 
 \begingroup
 \begin{table}[t]
\begin{ruledtabular}
\begin{tabular}{| c | c | c | c | c |}
  \hline
  Species & $\omega_{HFS}$~(GHz) & $\omega'_{HFS}$~(GHz) & $\Gamma$~(MHz) & $\chi$  \\
  \hline
  Na  & $1.722$ & $0.189$ & $9.8$ & $0.21$  \\
  $^{87}$Rb  & $6.834$ & $0.814$ & $5.75$ & $0.059$  \\
  $^{85}$Rb  & $3.036$ & $0.361$ & $5.75$ & $0.13$  \\
  Cs  & $9.193$ & $1.167$ & $4.57$ & $0.046$  \\
     \hline
  \end{tabular}
\end{ruledtabular}
  \caption{
  Values of $\omega_{HFS}$, $\omega'_{HFS}$, the linewidth $\Gamma$, and $\chi$~(evaluated for $\delta=0.5$ GHz) for different atomic species.
For Na and $^{87}$Rb, A$\equiv |F=2,m_F=-2\rangle$, and B$\equiv |F=1,m_F=-1\rangle$, and $\chi$ is defined as in Eq.~\eqref{eq:chi}. 
  For $^{85}$Rb,   A$\equiv |F=3,m_F=-3\rangle$, and B$\equiv |F=2,m_F=-2\rangle$, and $\chi=6\left [\frac{5}{27} \frac{\delta}{\delta+\omega_{HFS}}+
\frac{4}{27} \frac{\delta}{\delta+\omega_{HFS}-\omega'_{HFS}} \right ]$. 
For Cs, A$\equiv |F=4,m_F=-4\rangle$, and B$\equiv |F=3,m_F=-3\rangle$, and $\chi=6\left [ \frac{7}{48} \frac{\delta}{\delta+\omega_{HFS}}+
\frac{3}{16} \frac{\delta}{\delta+\omega_{HFS}-\omega'_{HFS}} \right ]$.
}
\label{table:T1}
\end{table}
 \endgroup

 
 \subsubsection{Bosonic case}

Let us first study the bosonic case. We consider the Raman scheme of Fig. 1 of the main text, where $L_{2,3}$ are $\sigma$ lasers, and $L_{1,4}$ are $\pi$ lasers. 
The lasers are characterized by their Rabi frequency $\Omega_{j=1,\dots, 4}$ proportional to the square root of their intensity, their frequency $\omega_j$, 
 and their wave vector ${\bf k}_j$. 
As mentioned above, we illustrate the case of bosonic species with the example of $^{87}$Rb, 
 with A$\equiv |F=2,m_F=-2\rangle$, and B$\equiv |F=1,m_F=-1\rangle$.  Let us consider in detail the different Raman-assisted hops we would like to 
 induce~(the Clebsch-Gordan factors may variate for other alkali species):
 \begin{itemize}
 \item (i) $(A,0)\to (0,A)$. This transition leads to an energy shift $\Delta E=-\Delta$. For $\omega_2-\omega_3=\omega_1-\omega_4=-\Delta$, the transition is realized 
by either $L_1$ and $L_4$, or by $L_2$ and $L_3$, with amplitude 
$V_{23}=\frac{1}{2}\frac{\Omega_2\Omega_3^*}{\delta}+\frac{\Omega_1\Omega_4^*}{\delta}\frac{1}{6}\chi$, with
\begin{equation}
\chi=6\left [\frac{1}{4} \frac{\delta}{\delta+\omega_{HFS}}+
\frac{1}{12} \frac{\delta}{\delta+\omega_{HFS}-\omega'_{HFS}} \right ].
\label{eq:chi}
\end{equation}
  \item (ii) $(A,A)\to (0,AB)$. This transition has $\Delta E=U_{AB}-\Delta$, and involves absorbing a $\sigma$ photon, and emitting a $\pi$ one. 
 We assume $\omega_2-\omega_4=U_{AB}-\Delta+U$, where $U$ is a small detuning~(within the Raman resonance width). 
 As discussed below this tunneling plays the role of effective on-site interactions in the final model.
The transition is quasi-resonantly provided by $L_2$ and $L_4$, with amplitude $V_{24}=\frac{1}{\sqrt{6}}\frac{\Omega_2\Omega_4^*}{\delta}$. 
  \item (iii) $(AB,0)\to (A,A)$. For this case, $\Delta E=U_{AB}-\Delta$, and is performed by absorbing a $\pi$ photon, and then emitting a $\sigma$ one. 
For $\omega_1-\omega_3=-U_{AB}-\Delta-U$, the transition is quasi-resonantly provided by 
$L_1$ and $L_3$, with Raman amplitude $V_{13}=\frac{1}{\sqrt{6}}\frac{\Omega_1\Omega_3^*}{\delta}$. 
 \item (iv) $(AB,A)\to (A,AB)$. For this transition $\Delta E=-\Delta$, and can be only realized 
 using $\pi$ lasers. For $\omega_1-\omega_4=-\Delta$, the transition is realized by  $L_{2,3}$, 
 with amplitude $V_{14}=\frac{1}{3}\frac{\Omega_1\Omega_4^*}{\delta}$.
 \end{itemize}
\noindent For  $\frac{|\Omega_1||\Omega_4|}{4}=\frac{|\Omega_2||\Omega_3|}{3}=
\frac{|\Omega_1||\Omega_3|}{2\sqrt{3}}=\frac{|\Omega_2||\Omega_4|}{2\sqrt{3}}=\Omega^2$, 
$\Omega_1=|\Omega_1|e^{-i\phi}$, and $\Omega_{j=2,3,4}=|\Omega_j|$, one obtains 
$V_{23}=\frac{\Omega^2}{\delta}(1+\chi e^{-i\phi})$, $V_{24}=\sqrt{2}\frac{\Omega^2}{\delta}$, 
$V_{13}=\sqrt{2}\frac{\Omega^2}{\delta}e^{-i\phi}$, and 
$V_{14}=2\frac{\Omega^2}{\delta}e^{-i\phi}$. The term proportional to $\chi$ in (i) is not discussed in the main text, for simplicity of the exposition. 
 It results in an additional term in the final Hamiltonian, as shown in Subsec.~\ref{subsec:Heff}.  However, $\chi$ can be kept small . Specific examples for different species for $\delta=0.2$ GHz are 
 shown in Table~\ref{table:T1}~\cite{SteckRb87,SteckRb85,SteckNa,SteckCs}. We show below that for small $\chi$ the spurious term does not modify the 
 conclusions of the 
 main text. 

 \subsubsection{Fermionic case}
  
For fermions it is better to choose A as the state with larger $F$, e.g. A$\equiv |F=9/2,-9/2\rangle$ and B$\equiv |F=7/2,-7/2\rangle$ for $^{40}$K. 
 In the scheme of Fig.~1 of the main text, we choose $L_{2,3}$ as $\pi$ lasers, and $L_{1,4}$ as $\sigma$ lasers. We may proceed as for bosons, but choosing $\Omega_1=-|\Omega_1|e^{-i\phi}$. 
 Due to 
 the inverted choice of A and B, we obtain: $V_{23}=\frac{\Omega^2}{\delta}$, $V_{24}=\sqrt{2}\frac{\Omega^2}{\delta}$, $V_{13}=\sqrt{2}\frac{\Omega^2}{\delta}e^{-i\phi}$, and 
$V_{14}=2\frac{\Omega^2}{\delta}(e^{-i\phi}+\chi) $. For $^{40}K$, 
$\chi=\left [ \frac{32}{243} \frac{\delta}{\delta+\omega_{HFS}}+
\frac{49}{243} \frac{\delta}{\delta+\omega_{HFS}-\omega'_{HFS}} \right ]$, with $\omega_{HFS}=1.286$ GHz, $\omega'_{HFS}=0.155$ GHz~(the linewidth is $\Gamma= 5.96$ MHz)~\cite{TieckeK}. 
For $\delta=0.2$ GHz, $\chi=0.072$. Note that inverting the choice of A and B, would lead to $V_{nm}$ of the same form as in the bosonic case but with $\chi$ four times larger ($\chi=0.28$  for $\delta=0.2$GHz). 
The latter justifies our choice of A and B.


\subsection{Effective Hamiltonian}
 \label{subsec:Heff}

We consider that only processes (i) to (iv)  are resonant, and postpone to the Subsec.~\ref{subsec:Spurious} the discussion about spurious processes.
We denote as $c_j$ the bosonic operator corresponding to the Fock-state manifold $\{ |0\rangle$, $|1\rangle\equiv |A\rangle$, $|2\rangle\equiv |AB\rangle\}$. 
Let us consider ${\bf k}_{1,2}=k{\bf e}_y$, and ${\bf k}_{3,4}=k{\mathbf e}_x$, and $k=\pi/D$. 

\subsubsection{Bosonic case}

For the bosonic case,  the Raman-assisted hopping of the previous subsection result in the effective 
Hamiltonian:

\begin{eqnarray}
H&=&-t \sum_j (-1)^j \left [ c_{j+1}^\dag e^{i\phi n_j} c_{j} +\mathrm{H.c.}  \right ] \nonumber \\
&+& \chi  \sum_j (-1)^j \left [ e^{i\phi} c_{j+1}^\dag (1-n_j )(1-n_{j+1}) c_{j} +\mathrm{H.c.}  \right ] \nonumber \\
&+& \frac{U}{2}\sum_j n_j(n_j-1).
\label{eq:Heff-1}
\end{eqnarray}
with $n_j=c_j^\dag c_j$, and 
$t=\left (\frac{\Omega^2/\delta}{2\Delta} \right ) J$.
Typical values of the Raman-assisted hopping rate, $t$, are of the order of few tens of Hz.
Note that the factor $(-1)^j$, which results from the $x$ projection of $\delta {\mathbf k}$, may be easily eliminated by redefining the bosonic operators in the form: 
$b_{4l}=c_{4l}$, $b_{4l+1}=c_{4l+1}$, $b_{4l+2}=-c_{4l+2}$, $b_{4l+3}=-c_{4l+3}$, with $l$ an integer. In this way we obtain:
\begin{eqnarray}
H&=&-t \sum_j \left [ b_{j+1}^\dag e^{i\phi n_j} b_{j} +\mathrm{H.c.}  \right ] \nonumber \\
&+& \chi  \sum_j \left [ e^{i\phi} b_{j+1}^\dag (1-n_j )(1-n_{j+1}) b_{j} +\mathrm{H.c.}  \right ] \nonumber \\
&+& \frac{U}{2}\sum_j n_j(n_j-1).
\label{eq:Heff-bosons}
\end{eqnarray}

\subsubsection{Fermionic case}

Proceeding in exactly the same way for the fermionic case, we obtain 
\begin{eqnarray}
H&=&-t \sum_j \left [ b_{j+1}^\dag e^{i\phi n_j} b_{j} +\chi b_{j+1}^\dag n_j n_{j+1} b_{j} +\mathrm{H.c.}  \right ] \nonumber \\
&+& \frac{U}{2}\sum_j n_j(n_j-1).
\label{eq:Heff-fermions}
\end{eqnarray}

\subsection{Effect of a finite $\chi>0$}

For $\chi = 0$, Hamiltonians~\eqref{eq:Heff-bosons} and~\eqref{eq:Heff-fermions} reduce 
to Eq.~(1) in the main text, and hence to
the anyon-Hubbard model (AHM). $\chi > 0$ results in spurious
collisionally-assisted hopping terms. However, as commented
above, $\chi$ can be kept small.

\begin{figure}[t]
\centering
\includegraphics[width=\columnwidth]{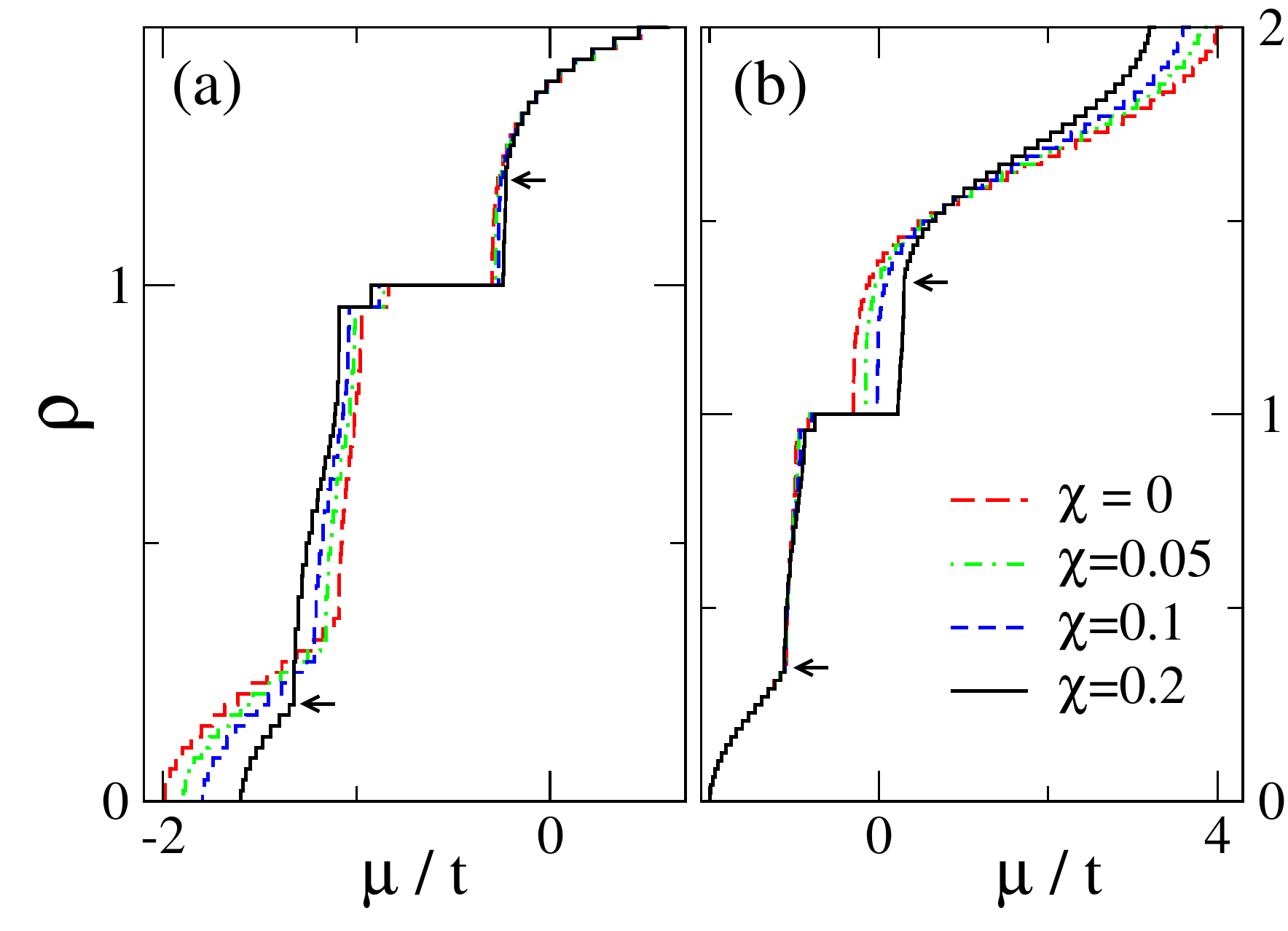}
\caption{Equation of state $\rho=\rho(\mu)$ of the AHM ($U=0$,
$\phi=\pi$, $L=48$) for different values of $\chi$ for the spurious
correlated tunneling in (a) Eq. (6) and (b) Eq. (7). Small arrows depict
the kinks in the $\rho(\mu)$ curve for $\chi=0.2$ characterizing the
PP-SF transition. }
\label{fig:S2}
\end{figure}

Figure~\ref{fig:S2}~(a) shows the effect of the spurious correlated
tunneling in Eq.~\eqref{eq:Heff-bosons} which basically effects the low-density part
$\rho<1$ and results in a small shift of the SF-PP transition
(indicated by small arrows) to lower $\rho$. Interestingly a PP phase may be observed also for
$\rho>1$ for sufficiently large $\chi$.

Figure~\ref{fig:S2}~(b) shows the effect of the spurious correlated
tunneling in Eq.~\eqref{eq:Heff-fermions}. The effect is negligible for small lattice
fillings $\rho < 1$. Due to its $\rho^2$ -dependence even for much
larger values of $\chi$ as discussed in the previous subsection the
curves of Fig.~\ref{fig:S2}~(b) basically overlap with the pure
AHM-case. For densities $\rho \geq 1$ the phase boundaries are slightly
modified. In particular the size of the Mott-insulator region increases
and the PP phase may now extend to a small region
above unit filling (see arrows in Fig.~\ref{fig:S2}~(b)).

Note that, interestingly, the choice of a larger detuning $\delta$ and thus larger
values of $\chi$ actually favors the observation of the PP phase in
experiments.



\subsection{Comparison to the model of Keilmann et al.}
 \label{subsec:Comparison}
 
Note that (i) and (iv) are energetically degenerated, but (apart from the small spurious effect discussed above) they may be addressed with different lasers due to selection rules. This point constitutes the major 
drawback of the proposal of Ref.~\cite{Keilmann2011}. In that proposal, a single component (A) was considered, and process (iv) was of the form (AA,A)$\to$(A,AA), which cannot be resolved 
from process (i). As a result, both $L_2$ and $L_3$, and $L_1$ and $L_4$ address both (i) and (iv), preventing the realization of the desired density-dependent Peierls phase. 
The two processes may be just discerned by considering a very small detuning $\delta\ll U_{AA},\Delta$. Since $U_{AA}$ and $\Delta$ are of the order of kHz, $\delta\lesssim 100$ Hz is necessary. 
However $\Gamma$ is typically of the order of few MHz (see e.g. Table~\ref{table:T1}), rendering unfeasible 
the proposal of Ref.~\cite{Keilmann2011}. 


 \begingroup
 \begin{table}[t]
\begin{ruledtabular}
\begin{tabular}{| c | c | c | c |}
  \hline
  Species &$a_{AA}/a_B$& $a_{AB}/a_B$ & $|\delta U/h|$~(kHz)  \\
  \hline
  Na  & $54.53$ & $64.5$ & $0.72$  \\
  $^{87}$Rb  & $100.36$ & $99.08$ & $0.23$  \\
  $^{85}$Rb  & $-469$ & $-386$ & $1.40$  \\
  Cs  & $-1980$ & $2436$ & $8.89$  \\
     \hline
  \end{tabular}
\end{ruledtabular}
  \caption{
  Values of $a_{AA}$, $a_{AB}$, and $\delta U/h$ for different bosonic species, for $s=20$ and $D=532$nm~(except for $^{87}$Rb where we employ $s=25$ and $D=266$nm)~\cite{Tiemann-Private}.
For Na and $^{87}$Rb, A$\equiv |F=2,m_F=-2\rangle$ and B$\equiv |F=1,m_F=-1\rangle$.
  For $^{85}$Rb,   A$\equiv |F=3,m_F=-3\rangle$, and B$\equiv |F=2,m_F=-2\rangle$. 
For Cs, A$\equiv |F=4,m_F=-4\rangle$, and B$\equiv |F=3,m_F=-3\rangle$. Note that negative scattering lengths are of no concern in our set-up. 
}
\label{table:T2}
\end{table}
 \endgroup



\subsection{Spurious processes}
 \label{subsec:Spurious}

Spurious Raman processes must be kept out of resonance:
\begin{itemize}
\item (v) (A,0) $\to$ (0,B): $\Delta E=-\Delta$
\item (vi) (A,A) $\to$ (0,AA): $\Delta E=-\Delta+U_{AA}$
\item (vii) (AA,0) $\to$ (A,A): $\Delta E=-\Delta-U_{AA}$
\item (viii) (AB,A) $\to$ (B,AA): $\Delta E=-\Delta+\delta U$, with $\delta U=(U_{AA}-U_{AB})$
\item (ix) (AA,B) $\to$ (A,AB): $\Delta E=-\Delta-\delta U$
\end{itemize}
Note that (vi), (vii), (viii) and (ix) are not possible for fermionic atoms. Process (v) is just possible with 
$J_{24}$ or $J_{13}$. But these laser combinations are (quasi-)resonant with $-\Delta\pm U_{AB}$. 
For a sufficiently large $U_{AB}$ process (v) is hence far from resonance with either  $J_{24}$ or $J_{13}$. 
As an illustration we consider the case of $^{40}$K atoms, with $a_{AB}\simeq 170 a_B$, with $a_B$ the Bohr radius, and a lattice with $D=532$ nm, and $s=20$, 
$U_{AB}/h\simeq 7$ kHz. 
Note that typical values for the width of the 
Raman resonance, $W\lesssim 50$ Hz~\cite{MiyakeThesis}. Hence $U_{AB}\gg W$, and hence process (v) may be safely 
ignored.

For bosons, (vi), (vii), (viii) and (ix) must be also considered. These processes can be only realized using  
$J_{23}$. To neglect the (vi) and (viii) processes one needs $U_{AA}\gg W$. This is the case in experiments, where typically $U_{AA}\sim 1$ to few $10$s kHz. In contrast, to avoid (viii) and (ix) one must 
demand $\delta U \gg W$. The latter condition is more strict, but may be attained as well in experiments. Examples for different bosonic species are illustrated in Table~\ref{table:T2}~\cite{Kronjager-Thesis,Tiemann-Private}.

\subsection{Initial preparation}
 \label{subsec:Initial}

If the spurious processes can be neglected, only on-site states 0, A, and AB may be reached by Raman-assisted hopping.  
However, B, AA, BB, and even sites with occupation higher than two, may initially exist. 
Sites with an occupation larger than two are eventually destroyed by three-body processes, and are hence of no major concern.
This is however not the case of B, AA and BB, which should be avoided in the initial preparation, since they would induce immobile spurious defects.
A straightforward way to avoid immobile defects is to consider a low-density gas of maximally singly occupied sites by A 
particles~(doubly occupied AA sites may be eliminated using photo association pulses~\cite{Sherson2010}). One may then compress this dilute gas increasing 
the density until reaching the desired density. Since A can just move into AB, doubly occupied sites will be guaranteed to have an AB occupation, 
and hence the on-site states remain always within the $\{$0, A, AB$\}$ manifold. \\



\begin{figure*}[ht]
\centering
\includegraphics[width=1.0\linewidth]{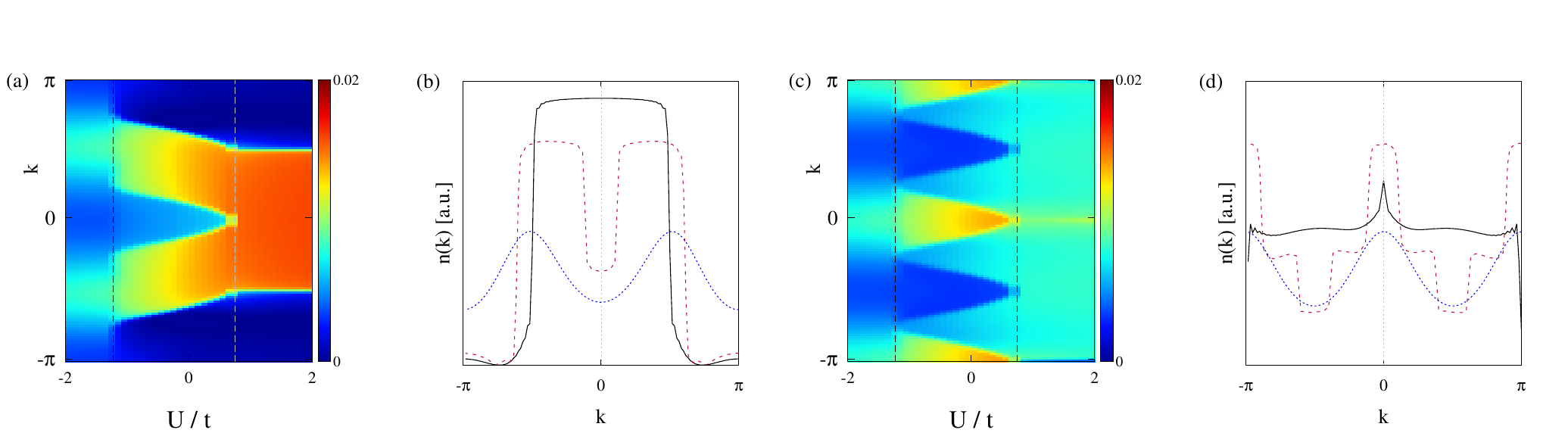}
\caption{Quasi-momentum distribution $n(k)$ of the A component for
the case of fermionic atoms as function of $U/t$
($\phi=\pi$, $\rho=1/2$, $L=60$ sites). (c) and (d) show $n(k)$
including the oscillating factors $(-1)^j$ created by the Raman-assisted
hopping; (a) and (b) show the same data, but excluding the oscillating terms. (b)
and (d) show $n(k)$ for $U=2$ (solid line), $U=0$ (dashed line) and
$U=-2$ (dotted line). In (a) and (c) the dashed lines denote the phase
transitions between PSF, PP and SF phases. }
\vspace*{-0.3cm}
\label{fig:S3}
\end{figure*}


\section{Quasi-momentum distribution}
 \label{sec:Quasimomentum}

As discussed above,  the AHM may be realized experimentally 
with both fermionic and bosonic species. While many
observables, most relevantly the density, do not depend on the particular
choice of constituents, others, in particular the momentum distribution, are not
gauge invariant. 

As shown in Fig.~4(b) and (c) of the main text, bosonic
species closely mimic the quasi-momentum distribution of the effective
model. As mentioned above, the oscillating factors $(-1)^j$ 
resulting from the necessary $x$ projection of $\delta {\mathbf k}$, may be easily eliminated by a gauge transformation. 
However, this transformation must be undone in order to recover the momentum distribution of components A and B. As a result,  
A and B components exhibit a doubled unit-cell, and the quasi-momentum distribution $n(k)$ is consequently trivially shifted and
duplicated. For clarity we only depict half the Brillouin zone for A and B in Fig.~4 of the main text.

It is interesting to depict as well the momentum distribution for the fermionic case.
Figures~\ref{fig:S3} (a) and (b), which depict $n(k)$ neglecting the $(-1)^j$ factors, show a
single Fermi-surface in the SF, two for the PP, and a
blurred $n(k)$ for the PSF. The oscillating $(-1)^j$ factors obscure
this simple interpretation
~(as shown in Figs. (c) and (d)). However PSF, PP and SF phases
may be still clearly discriminated from each other by their characteristic momentum
distribution function.

\bibliographystyle{prsty}